\documentclass[acmsigecom]{acmtrans2m}
\usepackage{graphicx} 
\usepackage[english]{babel}
\usepackage{amsmath}
\usepackage{amsfonts}
\usepackage{mathtools}
\usepackage{algorithm}
\usepackage{algpseudocode}
\usepackage{graphicx}
\usepackage{bbm}
\usepackage{xcolor}     

\usepackage[nameinlink]{cleveref}

\makeatletter
\newcommand\footnoteref[1]{\protected@xdef\@thefnmark{\ref{#1}}\@footnotemark}
\makeatother

\newtheorem{example}{Example}

\newcommand{\E}{\mathbb{E}}

\title{Recent Developments in Pandora's Box Problem: Variants and Applications}

\author[Hedyeh Beyhaghi and Linda Cai]{Hedyeh Beyhaghi \\ Carnegie Mellon University \and Linda Cai \\ Princeton University}

\begin{abstract} 
In 1979, Weitzman introduced Pandora's box problem as a framework for sequential search with costly inspections. Recently, there has been a surge of interest in Pandora's box problem, particularly among researchers working at the intersection of economics and computation. This survey provides an overview of the recent literature on Pandora's box problem, including its latest extensions and applications in areas such as market design, decision theory, and machine learning.
\end{abstract}

\begin{document}

\maketitle

\section{Introduction: the canonical Pandora's box problem} \label{sec:intro}
In many economic situations, search problems involve multiple options with unknown rewards. Gathering more information can reduce uncertainty about an option's reward but at a cost. The goal is to obtain a high-quality reward while minimizing the cost. For example, a company seeking to hire job candidates may need to conduct expensive onsite interviews to better assess candidate quality. Similarly, a student choosing between multiple university offers might need to visit campuses to gain a clearer understanding of their preferences for each institution.

The foundational model of optimal search, known as {\em Pandora's box problem}, was first established by~\citeN{Weitzman79}. The problem consists of a searcher who can choose a prize from one of the $n$ boxes. Each box contains a prize with an unknown value drawn from a distribution known to the searcher a priori. The value distributions of the boxes are independent from each other but may be different. The searcher can perform a sequence of actions, either opening a box or selecting a box. Opening box $i$ incurs a cost $c_i$, revealing the prize value $v_i$ inside, while selecting box $i$ yields a payoff of $v_i$ and terminates the search process. Importantly, the box must first be opened in order to be selected. 
The searcher devises an adaptive policy, which determines the next action based on previous actions and outcomes. The objective of the searcher is to maximize their expected utility, which is defined as the expected selected prize value minus the total inspection costs. As an illustration, we use the following running example by Weitzman.

\begin{example}[\cite{Weitzman79}]\label{ex:weitzman_two_boxes}
    Consider two boxes $A$ and $B$, where $A$ has a reward of $55$ or $100$ each with probability $0.5$ and a cost of $15$, and $B$ has a reward of $0$ or $240$, with probabilities $0.8$ and $0.2$, respectively, and a cost of $20$. 
\end{example}
Consider a potential strategy for the searcher as follows. The searcher opens box $B$ first. If the reward is $240$, the searcher selects the reward and terminates the search. If the reward is $0$, the searcher continues on to open box $A$ and takes the maximum reward observed (here the reward of box $A$). Using this strategy, for instance, when the reward of the first box is $0$ and the reward of the second box is $100$, the searcher opens both boxes, paying a total cost of $20 + 15 = 35$, and gains value $\max(0, 100) = 100$ resulting in a utility of $100 - 35 = 65$. Similarly, we can calculate the searcher's utility in other cases and use the probability of each case to find the expected utility,  which for this strategy is $78$. 

\subsection{Optimal Solution and Deferred-Value Interpretation}\label{sec:kww_reduction} 

At first sight, the solution space for Pandora's box problem seems extremely complicated. In fact, since the optimal policy could be fully adaptive (opening different boxes depending on the history of the boxes that have already been opened and their values), it is not even clear that the optimal policy for Pandora's box problem can be described in polynomial space as a function of the input size. 
Surprisingly, Weitzman proves that not only is the optimal solution to Pandora's box problem efficiently describable, it is highly structured.
Specifically, the optimal policy, named as {\em Pandora's rule} by Weitzman, is greedy and order non-adaptive (meaning that the inspection order of the boxes is determined apriori, although the policy can adaptively terminate the process). Given any box $i$, a \textit{reservation value} $\sigma_i$ can be computed based only on the prize value distribution and the cost for the particular box in question and is not dependent on the value or cost of other boxes.\footnote{The reservation value $\sigma_i$ solves the equation $c_i = \E[\max(v_i-\sigma_i,0)]$,
	where the expectation is over the value distribution of box $i$. The reservation value turns out to be a  special case of indices proposed by \citeN{Gittins79}. See \citeN{Gittins79}, \citeN{Weber92}, and \citeN{Gittins11} for more detail on the Gittins index.} The optimal policy orders the boxes by nonincreasing reservation value and selects the largest observed value once this value exceeds the reservation values of all remaining boxes. Going back to \Cref{ex:weitzman_two_boxes}, as Weitzman shows, although box $A$ has a higher expected value, lower cost, higher minimum value, and lower variance and may seem a better option to try first, surprisingly, it has a lower reservation value and box $B$ will be the first box to open in the optimal solution. Intuitively, by opening box $B$ first, the searcher gains more information about future actions -- one can verify that if the searcher opens box $A$ first, the next best action is to open box $B$ regardless of the observed value from box $A$.

Almost forty years after the introduction of the problem and its optimal solution, \citeN{KleinbergWW16} provide a new interpretation of Pandora's rule that opens the path to new directions in understanding search problems with cost. While Weitzman uses a local improvement argument to prove the optimality of Pandora's rule, Kleinberg et al.\ reduce Pandora's box problem to a related search problem where the values of items are revealed for free. Specifically, Kleinberg et al.\ define the \textit{deferred-value} of a box $i$ as the minimum between the prize value $v_i$ and the reservation value $\sigma_i$. They then prove that the expected maximum deferred value upper bounds the utility of any policy for Pandora's box problem. Finally, using specific structural properties of Pandora's rule, they show that the expected utility from Pandora's rule is exactly the expected maximum deferred value.\footnote{The reduction of Kleinberg et al.\ is closely related to the idea of ``prevailing charges" in \citeN{Weber92}'s proof of the Gittin's index theorem. The deferred value reduction of Kleinberg et al.\ has antecedents in ~\citeN{ArmstrongV15} and has been independently discovered by~\citeN{Armstrong17} and~\citeN{ChoiDK18}.}
As an illustration, in \Cref{ex:weitzman_two_boxes}, the reservation value for boxes $A$ and $B$ are $70$ and $140$ respectively, therefore, the deferred value distribution of box $A$ is $55$ or $70$, each with probability $0.5$ and $B$ is $0$ with probability $0.8$ and $140$ with probability $0.2$. The expected maximum of these deferred value distributions with no cost is $0.2 \times 140 + 0.8 (0.5 \times 55 + 0.5 \times 70) = 78$, which is exactly equal to the maximum expected utility in \Cref{ex:weitzman_two_boxes}. \\

\noindent Since the re-introduction of Pandora's box problem by Kleinberg et al.\ to the theoretical computer science community and especially the community working at the intersection of economics and computation, many variants, extensions, and applications of the original model have been considered. For the rest of the survey, we summarize these directions and highlight some of the common themes and techniques.

\section{Variants and Extensions} \label{sec:variant}

In this section, we overview variants and extensions of Pandora's box problem that have been considered in recent literature. Motivated by the characteristics of specific search applications, these variants and extensions either relax or restrict key aspects of the original model. For instance, there may be multiple ways to inspect a box (\Cref{sec:variant:alt}), the searcher's value for the boxes can be correlated (\Cref{sec:variant:correlated}), the cost of inspection may not be additive (\Cref{sec:variant:cost}), and the searcher may be able to choose more than one item (applied across many variants, often combined with other modifications to the model). In terms of restrictions, the searcher may not be able to inspect in any order they want (\Cref{sec:variant:order}), they may not be able to select a previously inspected option that they passed on (\Cref{sec:variant:norecall}), and they may not have exact knowledge about their value distribution for a box (\Cref{sec:variant:correlated}).

Before diving into the specific variants, we will first discuss variations on the objective function and solution concepts that will be used throughout the rest of the section. Firstly, the objective function of the search problems can either be formulated as utility maximization (each box has a non-negative value and the goal is to maximize the selected value minus total cost) or loss minimization (each box has a non-negative price and the goal is to minimize the selected price plus total cost), and both objectives have been studied since the inception of the optimal search problem \cite[Chapter~3]{Degroot70}. For exact optimization, the two objectives are equivalent; however, approximating the optimal loss is often easier than approximating the optimal utility. When discussing the variants, we consider the more commonly used utility maximization objective as the default objective, except when otherwise explicitly stated.

In terms of solution concepts, it may not be possible to efficiently describe or compute the optimal policy among all possible policies for certain variants. To overcome this, it is helpful to focus on more limited classes of policies with better descriptive or computational properties. The three most commonly considered solution concepts are as follows:

\begin{itemize}
    \item\,
    \textbf{Fully adaptive policy}: the most general class of policies, where the action of the policy can depend on previous actions and the values it has seen. 
    \item\,
    \textbf{Order non-adaptive policy}: the class of policies where the inspection order of the boxes is predetermined before the value of any box is revealed. However, the stopping rule, i.e., when the policy terminates the process, may be adaptive. 
    \item\,
    \textbf{Fully non-adaptive policy}: the class of policies where both the inspection order and the stopping rule are non-adaptive. In particular, the policy would always inspect all boxes that are specified in the inspection order. 
\end{itemize}

\subsection{Alternative Inspection Methods} \label{sec:variant:alt} 
In many search applications such as student choosing universities or consumer search, there may be several different inspection methods (e.g., online research vs.\ visiting in person), and inspection may not be required before selecting a box. Models in this subsection relax the original model to allow such variations in methods of inspection.

The most well-studied thread under this relaxation is the {\em nonobligatory} inspection model. In this model, instead of having to inspect before selecting a box, the searcher can alternatively claim the box closed without inspection and get the expected value of the box. This model has been introduced independently in different communities (wireless network, stochastic testing, search theory) and under different names \cite{GuhaMS08,ChangL09,AttiasKLS17,Doval18}. In particular, \citeN{Doval18} formulated the nonobligatory inspection model explicitly as a generalization to the original Weitzman's Pandora's box problem and popularized the model in the economics and computation community. Recently, a steady line of work \cite{GuhaMS08,Doval18,BeyhaghiK19,FuLL23,BeyhaghiC23} resolved both the computational complexity and approximability of the problem. 

The literatures on complexity, structure and approximability of the non-obligatory inspection model progressed in conjunction and are deeply intertwined. \citeN{GuhaMS08}\footnote{\citeN{GuhaMS08} studied the problem under the context of stochastic probing in wireless networks. The wider community was unaware of their work until recently.} first show a significant structural result: the optimal policy claims a unique box closed across all decision branches. Using their structural result, Guha et al.\ show that the competitive ratio of \textit{committing policies} (order non-adaptive policies where the searcher commits ahead of time to whether they will inspect each box prior to selecting it) is exactly $0.8$. Independently, \citeN{BeyhaghiK19} show that the competitive ratio of committing policies is at least $1 - 1/e \approx 0.63$ by a reduction to stochastic submodular maximization, which can also be applied to more general models as we will discuss later. 

\citeN{Doval18} provides evidence both for the complexity of the optimal policy and the existence of additional structure. In particular, Doval shows that the optimal policy may be order-adaptive, while showing that the optimal policy has a two-phased structure (where the inspection order only changes once) under additional assumption on the value distribution.\footnote{Specifically, \citeN{Doval18} considers the binary prize environment, where the value of each box $i$ is supported on $\{L, H_i\}$, where the low value in the support is shared between all boxes, but the high value in the support may be distinct for each box.}

Finally, \citeN{FuLL23} prove that the problem is NP-hard, which confirms intuition in previous literature. In terms of structural results, \citeN{FuLL23} and \citeN{BeyhaghiC23} show that in the \textit{general} setting, the optimal policy is two-phased and can be fully specified through an initial inspection order and a threshold for each box.\footnote{The results in \citeN{FuLL23} and \citeN{BeyhaghiC23} build on the structural results in \citeN{GuhaMS08} and \cite{Doval18}.} As a consequence, the decision version of the problem is in NP (since one can prove the optimal utility is above a certain threshold by succinctly describing the policy that obtains this utility). Further, Fu et al. and Beyhaghi and Cai provide a PTAS for the nonobligatory inspection model. 

In the nonobligatory inspection model, there are two ways of inspecting a box: pay the full cost and inspect the box or claim the box closed without inspection. However, there may be other options that lie in between: perhaps a smaller cost is needed to reduce the variance of the value distribution. The remaining variants in this subsection addresses the ``in between" inspections. \citeN{KleinbergWW16} consider an alternative model where there are multiple stages of inspection, and the searcher could only claim the value in the box after all stages of inspections are completed. As the searcher progresses through the stages, more information about the value is revealed, and more cost is incurred. The searcher can stop examining the model further at any stage. Kleinberg et al.\ find that the optimal policy for this staged inspection setting is a generalized form of Pandora's rule, where a reservation value can be computed for a box at any stage, and the searcher always inspects the box with the highest reservation value (given its current stage). 

In a similar spirit, \citeN{keV19} consider a model where information is revealed gradually, but the searcher can claim the box (or stop) at any point. In addition, in their model, the discovery process is continuous rather than discrete, and the value of each box is binary supported.\footnote{Note that the support for different boxes could be different, but the value of each box only has two possibilities.} They find that even in the case when there are two boxes and a fixed-valued outside option, the solution space may be complicated; they characterize the optimal policy under conditions such as when the outside option is below or above certain thresholds.

\citeN{AouadJS20} introduce a model where the searcher has two ways of opening a box: fully open and partially open. Similar to \citeN{KleinbergWW16}, a box must be fully opened before the value can be claimed. However, unlike the model in Kleinberg et al., the searcher can fully open a box without partially opening the box first. Aouad et al.\ prove that the best \textit{committing policy}
\footnote{ \label{committingPolicy}As defined in our discussion of the non-obligatory inspection model.}
 is $(1-1/e)$-competitive to the optimal utility using an analysis inspired by \citeN{BeyhaghiK19}. Moreover, they show that any committing policy or its negation (flipping which box should be partially opened versus not partially opened) is $1/2$-competitive to the optimal utility. Aouad at al.\ also design a simple threshold-based committing policy that is near optimal when the number of items is sufficiently large. 

Finally, as a direct extension to the non-obligatory inspection model, \citeN{Beyhaghi19} introduces {\em Pandora's box problem with alternate inspection} model, where the searcher has $k$ different methods for inspecting each box (including not inspecting at all), and the searcher can select at most one method for each box. \cite{BeyhaghiK19,Beyhaghi19} prove that \textit{committing policies}\footnoteref{committingPolicy} are $(1 - 1/e)$-competitive in this model.

\subsection{Search Without Recall} \label{sec:variant:norecall}
Even before Weitzman's seminal paper, the economics literature studied both the optimal search problem with recall (the searcher can select alternatives that they have seen in the past) and without recall (the searcher has to select the item or forgo it forever) \cite[Chapter~3]{Degroot70}. Hybrid settings where recall is uncertain have also been considered \cite{karniS77}. In modern literature, the version of Weitzman's Pandora's box problem without recall is studied under the name {\em Committed Pandora's box} \cite{FuLX18}. 

\citeN{KleinbergWW16} first showed that a simple threshold-based policy for Committed Pandora's box with arbitrary (possibly adversarial) constraints on inspection order achieves at least $1/2$ of the optimal utility by a reduction to prophet inequalities (an online selection model that is similar to Committed Pandora's box, but without inspection costs). The connection between (variants of) Committed Pandora's box and (variants of) prophet inequalities is repeatedly exploited by subsequent work.\footnote{\label{noteProphet} See \citeN{Lucier17} and \citeN{CorreaFHOV19} for in-depth discussions on prophet inequalities and variants.} 

\citeN{FuLX18} and \citeN{Segev021} provide a general framework for deriving polynomial time approximation schemes (PTAS) and efficient polynomial time approximation schemes (EPTAS) for stochastic optimization problems, respectively. As an application of their framework, Fu et al.\ show that Committed Pandora's box problem with free order selection (the searcher has the full freedom to pick the inspection order) has a PTAS by a direct reduction.  Segev and Singla show that, in fact, Committed Pandora's box problem has an EPTAS by first reducing the problem to free order prophet inequality,\footnoteref{noteProphet} which has an order non-adaptive optimal policy, and then applying their framework. 

\citeN{EsfandiariHLM19} study Committed Pandora's box problem under adversarial order and where the searcher is allowed to collect multiple prizes subject to general feasibility constraints (e.g., cardinality, knapsack or matroids constraints). In addition, in their model, the prize values and costs are drawn from a joint distribution, and the cost is only revealed after opening the box.
Esfandiari et al.\ prove that all variants of Committed Pandora's box problem they consider can be reduced to a corresponding prophet inequality problem with known competitive ratios.
Further, they extend Committed Pandora's box with adversarial order (with the objective of selecting one prize per round) to the contextual bandit setting and obtain a $1/2$-competitive policy based on the reduction to prophet inequalities.

\subsection{Restricted Order of Inspection} \label{sec:variant:order}
As we have discussed in \Cref{sec:variant:norecall}, both free order (no restriction) and adversarial order (complete restriction) are standard assumptions for online selection problems and have been considered in the context of Pandora's box problem. Motivated by applications such as funding research development, a more general question can be asked: what if there are \textit{some} restrictions on the searcher's inspection order? 

\citeN{BoodaghiansFLL20} initiate the study of partial order constraints such as tree or DAG like order restrictions for Pandora's box problem. In the case of (single selection) Pandora's box problem with tree or forest like order constraints, where a box can only be opened once its parent box is opened, Boodaghians et al.\ show that an order-dependent version of the reservation value
can be computed for each box, and the optimal policy always opens the remaining box (if any) with the largest reservation value.\footnote{Interestingly, the authors mention that Pandora's box problem with forest like order constraint is closely related to the branching bandit process studied in \cite{weiss88,KellerO03}, and their optimal policy has a similar structure.}  On the other hand, for variants where multiple selections are allowed (e.g., under matroid feasibility constraints or more general constraints), or when the order constraint is DAG like (where a box can be opened only if one of its predecessor boxes is opened), Boodaghians et al.\ show that finding a fully adaptive policy that achieves $\epsilon$ fraction of the optimal utility is NP-hard. They also propose a relaxed definition of approximately optimal policies and analyze the adaptivity gap between fully adaptive policies and fully non-adaptive policies.

\subsection{Beyond Independence and Distribution Assumptions} \label{sec:variant:correlated}

This section considers relaxations on two of the constraints in the original model: the independence of the searcher's valuation among boxes, and the knowledge of the distributions (or even having sample access to the distributions). Since \citeN{CGTTZ20} introduced both relaxations in the same paper, the study of these variations is often interweaved and thus presented here in a single section. Depending on whether the valuations are independent or correlated and whether historical samples are available (distributional learning setting) or the search is a repeated process without historical samples (online learning setting), there are four different variants that are discussed in this section.

\citeN{CGTTZ20} study the correlated value and distributional learning setting, mainly under the loss minimizing objective. Specifically, the prices in the boxes are drawn from an arbitrarily correlated joint distribution; moreover, the searcher is limited to $\mathsf{poly}(n)$ samples from the joint distribution, where $n$ is the number of boxes. 
Chawla et al.\ show that approximating the loss of the optimal fully adaptive policy within any sublinear factor requires exponential samples. Then, they efficiently find an order non-adaptive policy that is a constant approximation to the best order non-adaptive policy. Moreover, unless $\mathsf{P}=\mathsf{NP}$, one cannot efficiently find a fully adaptive policy that exceeds the best order non-adaptive policy. Chawla et al.\ also show that if their model has the utility maximization objective instead, no computationally efficient fully adaptive policy can even be a constant approximation to the best order non-adaptive policy. As a direct follow-up to Chawla et al.,
\citeN{GergaETC23} show that a generalized version of Weitzman's policy is constant-competitive against the best order non-adaptive policy. Compared to the linear programming rounding approach in Chawla et al., Gergatsouli and Tzamos's construction of a competitive order non-adaptive policy is more explicit while obtaining an improved competitive ratio. 

\citeN{GergatsouliT22} study the correlated value model under the more restrictive setting of online learning and show that when the prices of the boxes are selected by an oblivious adversary in each round, results of Chawla et al.\ extend to both the full information setting (values of all the boxes are revealed after each round) and bandit setting (only values of the boxes that the policy opened are observable). 

\citeN{GuoHTZ21} study the independent valuation and distributional learning setting, and prove that the searcher could obtain an $\epsilon$-additive approximation to the optimal utility with high probability given $\tilde{O}(\frac{n^3}{\epsilon^3})$ samples.~\citeN{FuL20}\footnote{Although published out of order, \citeN{GuoHTZ21} preceeds \citeN{FuL20} and is cited as prior work in all versions of \citeN{FuL20}. } improve this sample complexity to $\tilde{O}(\frac{n}{\epsilon^2})$.                               
Finally, \citeN{ACGPT22} and \citeN{GatmiryKSW22} study the independent valuation and online learning setting. Atsidakou et al.\ extend the original Pandora's box model (with loss minimizing objective) to the contextual bandit setting, where each round comes with potentially different sets of boxes. At the beginning of each round, the context and cost of the boxes are revealed up front, while the distribution of values in the boxes remains unknown (and can be different from round to round). Atsidakou et al.\ prove that as long as the context can be used to estimate the reservation value of the box, sub-linear regret against the optimal policy with full distributional knowledge (i.e., Pandora's rule) is achievable for both the full information and bandit setting. 
\citeN{GatmiryKSW22} study the online Pandora's box problem (with utility maximizing objective) where the boxes have unchanging prize value distributions that are unknown to the searcher. The value distributions are independent but not necessarily identical. Gatmiry et al.\ prove that in the bandit setting, the searcher can achieve $O(\mathsf{poly}(n) \sqrt{T})$ regret by an algorithm that estimates and then shrinks a confidence interval on each box's reservation value, where $T$ is the number of time steps. 

\subsection{Beyond Additive Costs}  \label{sec:variant:cost}
For many applications that motivate Pandora's box problem, such as students selecting universities and job search, the cost of inspection may not be additive. For instance, students who visit universities in nearby locations back to back may experience lower costs compared to having three separate trips to those universities. ~\citeN{BergerEFF23} extend Weitzman's Pandora's box problem by considering more general classes of cost functions such as submodular, XOS, or sub-additive functions. Their main result shows that the optimal policy for Pandora's box problem is order non-adaptive for submodular cost functions. On the other hand, when the cost function is XOS or sub-additive, adaptivity is required for the optimal policy. They also show that even for the more restrictive class of submodular cost functions, approximating the utility of Pandora's box problem requires an exponential number of queries to the cost function. 

\section{Applications} \label{sec:app}
In this section, we overview applications of Pandora's box problem in 
combinatorial optimization (\Cref{sec:app:opt}), mechanism design (\Cref{sec:app:mech}), delegation (\Cref{sec:app:delegate}) and matching markets (\Cref{sec:app:match}). The elegant structure of optimal or approximately-optimal solutions to Pandora's box problem plays a crucial role in addressing domain-specific problems where information acquisition is costly.  

\subsection{Combinatorial Optimization} \label{sec:app:opt}
\citeN{KleinbergWW16} and~\citeN{Singla18} 
applied the structures and tools from Pandora's box problem in a wider range of combinatorial optimization problems, such as the costly counterpart of maximum weighted matching, maximum knapsack, minimum vertex cover, minimum set cover, minimum facility location, and minimum prize-collecting Steiner tree. Kleinberg et al.\ initiated this thread by 
applying their no-cost reduction explained in \Cref{sec:kww_reduction}. Later,~\citeN{Singla18} expanded upon this reduction technique and applied it to a broader range of problems. Singla provides a general transformation for converting {\em frugal algorithms} (a type of greedy algorithm) into policies for solving combinatorial counterparts of Pandora's box problem, where the searcher can choose multiple boxes subject to feasibility constraints on the selected set. This transformation applies in both utility maximization and loss minimization settings.

\subsection{Mechanism Design} \label{sec:app:mech}

We overview several mechanism design papers with costly information acquisition that utilize
the structure of optimal or approximately optimal solutions to
 Pandora's box problem. These papers consider a few different scenarios between sellers and buyers, with a costly investigation process on one side of the market or the other.

\citeN{CremerSZ07} consider an auction scenario for selling a single item, where the set of buyers is not exogenous or determined in advance, and the seller needs to go through a costly sequential process to inform potential buyers about the auction. They show that in the case of independent buyers' valuations, the seller's problem can be reduced to Weitzman's model, where the distribution of each box is the Myerson virtual value distribution for each buyer.

\citeN{KleinbergWW16}, also focus primarily on the sale of an item to buyers. However, unlike the previous scenario, the buyers are informed about the auction but need to incur inspection costs to determine their values. Using their reduction from costly information acquisition to optimization with no cost (discussed in \Cref{sec:kww_reduction}) as a key element, they devise a descending price auction that achieves the same efficiency as a first price auction with modified value distributions but no cost of inspection, resulting in a small price of anarchy (approximate optimality). 
Later, \citeN{AlaeiMM21} extend the revenue maximization setting of Kleinberg et al.\ to the nonobligatory inspection model. They provide mechanisms, both for selling a single item and multiple copies of an item, that are approximately optimal even when the buyers arrive in an adversarial order. 
Subsequently, \citeN{WuJC22} also consider revenue maximization in a nonobligatory inspection setting; however, they particularly focus on the role of bundling the items in optimizing revenue. They study two different markets; the first with one mature and one new product, and the other with two new products. The valuation uncertainty only exists for new products. They show that in a market with one mature and one new product, bundling encourages search, while in a market with two new products, it discourages search. Orthogonally,~\citeN{FuL20} use the correspondence between the descending price auction in a costly information acquisition setting and the first price auction in the classic setting, developed by Kleinberg et al., to provide sample complexity bounds for auction design with costly inspections.

\citeN{Armstrong17} examines a scenario in which buyers intend to purchase a product, such as a book from an online marketplace, from one of several available sellers. A buyer learns their value and the price of a product upon inspection unless the seller advertises their price, in which case the price is known a priori. When a buyer purchases a product, they get utility equal to their value for the product minus the price and inspection cost set by the seller. From a buyer's perspective, their search problem is exactly equivalent to the canonical Pandora's box problem, and they are modeled to be employing Pandora's rule. Consequently, the reduction of \citeN{KleinbergWW16} and \citeN{ArmstrongV15} also applies to Armstrong's setting with prices and can be used to calculate a buyer's expected utility. Armstrong instead focuses on analyzing the seller's strategy decisions (setting product prices, using advertising to guide consumer searches, and determining the consumer's search costs) and their implications, 
both when a monopolist seller owns multiple products and when there are multiple sellers. Armstrong presents a detailed discussion of the factors that influence which sellers raise or lower their prices given the buyer's inspection order. In addition, Armstrong explores why it might be profitable for a seller to {\em obfuscate} the searcher by increasing its own inspection cost and examines the equilibria of the buyer-seller optimization problem. 

\citeN{ChoiDK18} examine a pricing game where a group of sellers with substitutable items compete with each other while buyers have partial information about their values. In the first step, the sellers simultaneously announce their prices. In the second step, the buyers go through a costly search process among sellers depending on the announced prices and their partial information about their values. The authors characterize buyers' optimal behavior and analyze the pricing game among the sellers.

\citeN{ChenBD22} propose a three-step mechanism for manufacturers outsourcing their production to suppliers to reduce procurement costs. First, suppliers submit price bids for contracts. Second, buyers investigate ways to reduce production costs, subject to a limit on the number of investigations. Third, the buyer awards the contract to the supplier with the lowest updated bid. The second step, which is a costly investigation process, is equivalent to a variant of Pandora's box problem where there is a limit on the number of boxes that can be opened. Although Weitzman shows that, generally, Pandora's rule may not be optimal given the limitation on the number of inspections, Chen et al.\ identify sufficient conditions for Pandora's rule to be optimal for buyer investigation, in which case the structural properties of Pandora's rule can be used to design the optimal three-step mechanism.

\subsection{Delegated Search} \label{sec:app:delegate}

Delegation in search problems refers to the process of a {\em principal} assigning a search problem to an {\em agent}, who possesses the necessary resources but may have interests that differ from those of the principal. A key question when considering a delegated search problem is how much the principal loses when they delegate the search to an agent. This quantity is referred to as the {\em delegation gap}. Although the theory of delegation was introduced in economics much earlier by the work of~\citeN{Holmstrom78} and~\citeN{Holmstrom84}, {one of the early papers that use ideas from Pandora's box problem to design optimal delegated search mechanisms is by \citeN{Postl04} who establishes a condition that ensures there is no loss in the delegation for the same-cost two-box version of the problem. Later,~\citeN{KleinbergK18} use a model introduced in~\citeN{ArmstrongV10} and incorporate ideas from Pandora's box problem to create nearly optimal delegated search mechanisms. One of the models they study is a costly information acquisition delegated search problem with binary options. This model involves a set of options, each with a probability of being feasible for the principal and a cost to investigate. They design a search mechanism with a limited delegation gap.~\citeN{BechtelDP22} later expand on the binary case to include matroid feasibility constraints. However, they prove that there is no constant-factor delegation gap beyond the binary model. To overcome this, they explore other variations, such as the shared-cost model, where the principal can choose how to split the costs with the agent before the delegation. They demonstrate that the shared-cost model has a constant-factor delegation gap for specific constraints.

\subsection{Matching Markets}\label{sec:app:match}

\citeN{ImmorlicaLLL20} explore a generalization of Pandora's box model within the context of matching markets, specifically focusing on many-to-one markets such as student-college mappings. In these scenarios, students must undergo a costly information acquisition process to determine their values for each college, while colleges maintain a publicly known ranking system for students. The authors introduce {\em regret-free stability} as a refined solution concept that builds upon the traditional {\em stability} definition in matching market literature, ensuring optimal information acquisition for students, and they demonstrate the existence of such a solution.

In a single-student model, the problem simplifies to the original Pandora's box problem, making Pandora's rule the optimal solution for the student's search. However, when multiple students are involved, the available college options for each student depend on the valuations of their peers. This interdependency between students' information acquisition choices is resolved by using approximate cutoffs (i.e., the lowest admissible student ranking for each college). With these cutoffs, students can independently tackle Pandora's box problem for the set of colleges where their rank meets the cutoff, ultimately achieving regret-free stability.

\section{Discussion: Common Structural and Technical Themes}

Although the variants and applications discussed in \Cref{sec:variant,sec:app} often extend or utilize Pandora's box problem in orthogonal directions, several concepts and ideas appear to be relevant across numerous variants and applications.

In the study of Pandora's box problems, a common theme is analyzing the relative power of simple policies, which typically refers to order non-adaptive policies, and comparing them to fully adaptive ones. This comparison is similar to the concept of the {\em adaptivity gap} in combinatorial optimization. The variants discussed in~\Cref{sec:variant} can be categorized into three classes: order non-adaptive policies being as powerful as fully adaptive policies, having a constant competitive ratio, and exhibiting a super-constant gap between them.
The original Pandora's box problem and some models with restricted order of inspection (\Cref{sec:variant:order}), beyond additive cost (\Cref{sec:variant:cost}), and search without recall (\Cref{sec:variant:norecall}) possess optimal order non-adaptive policies, placing them in the first class. In contrast, various models involving alternative inspection methods (\Cref{sec:variant:alt}) may not have optimal order non-adaptive policies, but they are constant-competitive against the best fully adaptive policy, falling into the second category. Additionally, the optimal policy in these cases may require limited adaptivity.
More sophisticated models with restricted order of inspection (\Cref{sec:variant:order}) and correlated distribution (\Cref{sec:variant:correlated}) belong to the third class, as they exhibit a super-constant utility gap between order non-adaptive and fully adaptive policies. 

In essence, for different Pandora's box problem variants, the adaptivity gap serves as an indicator of the problem's structural complexity. A large adaptivity gap, combined with an impossibility result in approximating the optimal fully adaptive policy, can motivate researchers to focus on approximating the optimal order non-adaptive or fully non-adaptive policy instead.

Another common technical theme is the reduction of a variant of Pandora's box problem with cost to a related problem without cost. This reduction is most direct in Pandora's box problem without recall setting (\Cref{sec:variant:norecall}), where many variants can be reduced to different variants of the prophet inequality problem. In the latter problem, the searcher's values for the boxes are drawn from known distributions and must select a box without recall, but revealing the value does not come at a cost. For the original Pandora's box problem with recall,~\citeN{KleinbergWW16} first used a cost-to-no-cost reduction in their alternative proof of optimality for Pandora's rule. Interestingly, a similar reduction can also be applied to Pandora's box problem with alternative inspection methods (\Cref{sec:variant:alt}) and the more general combinatorial optimization setting (\Cref{sec:app:opt}). The presence of cost in Pandora's box problem complicates the design of optimal or approximately optimal policies (approximating values and costs separately may not lead to approximately-optimal utility). Consequently, the reduction from cost to no cost often serves as a useful tool in revealing the structure of the optimal policy or facilitating the design of an approximately optimal policy. 

\section{Conclusion and future directions} 

The recent growing body of research on Pandora's box problem, presented in this survey paper, has introduced numerous variations and applications, indicating substantial potential for future exploration and investigation.

The alternative and generalized models of Pandora's box overviewed in \Cref{sec:variant} are far from exhaustive (especially those related to order restriction (\Cref{sec:variant:order}), correlated value distributions (\Cref{sec:variant:correlated}), and non-additive cost functions (\Cref{sec:variant:cost})), and warrant further investigation. In addition, the relationship between Pandora's box variants and broader stochastic optimization problems (e.g., Markov chains, multi-armed bandits) has been noted in multiple studies and can benefit from a systematic analysis. 
We also anticipate the emergence of future models that deviate from existing variants, exploring different aspects of the searcher's decision-making. For example, current models do not incorporate behavioral economic findings, such as risk and loss aversion. Empirical evidence from \citeN{bhatiaHZW21} demonstrates that these factors align more closely with human decision-making behavior in costly information acquisition settings. In another potential variant, the searcher may have a long time horizon with the boxes emerging and disappearing in an online fashion. Alternatively, there may be random signals from global events (e.g., the emergence of new technology) that provide information for all or a large segment of the boxes (e.g., a readjustment to the skills of applicants or their distribution). 

In the application domain, Pandora's box model is relevant to most applications where the cost of information acquisition is significant, including those that are not mentioned in \Cref{sec:app}, such as voting and advertisement. In voting scenarios, both the candidates and the voters may engage in costly investigations; e.g., the candidates optimize their investment of targeted campaigning to select populations, while the voters may face a similar trade-off as in canonical examples of Pandora's box model, where they choose between selecting well-known candidates and investigating their preference alignment with the less well-known ones. Similarly, in advertisement applications, the advertisers go through a costly investigation to select what populations to target and what advertisement methods to use. Other potential applications include models of labor, product, and financial markets with costly information acquisition. In addition, most existing applications employ the canonical Pandora's box model. Recent work on the variants proposed in \Cref{sec:variant} offers a wider range of modeling choices, and may enable greater realism for some applications.

\section{Acknowledgement}
The authors are grateful to Laura Doval, Brendan Lucier, and the editors-in-chiefs Irene Lo and Sam Taggart for their valuable feedback on earlier versions of the survey and identification of missing relevant work.

\bibliographystyle{acmtrans}
\bibliography{ref}

\end{document}